\RequirePackage{lineno}
\documentclass[prb,twocolumn,showpacs,amsbsy,linenumbers,floatfix,superscriptaddress]{revtex4}
\usepackage{epsfig,color}
\usepackage[cp1250]{inputenc}
\usepackage{amsmath}
\begin{document}

%\linenumbers
\title{Fano resonances and electron spin transport\\ through a two-dimensional spin-orbit-coupled quantum ring}

\author{M. P. Nowak}
\affiliation{AGH University of Science and Technology, Faculty of Physics and Applied Computer Science,\\
al. Mickiewicza 30, 30-059 Krak\'ow, Poland}
\affiliation{Departement Fysica, Universiteit Antwerpen, Groenenborgerlaan 171,
  B-2020 Antwerpen, Belgium}
\author{B. Szafran}
\affiliation{AGH University of Science and Technology, Faculty of Physics and Applied Computer Science,\\
al. Mickiewicza 30, 30-059 Krak\'ow, Poland}
\author{F. M. Peeters}
\affiliation{Departement Fysica, Universiteit Antwerpen, Groenenborgerlaan 171,
  B-2020 Antwerpen, Belgium}

\date{\today}

\begin{abstract}
Electron transport through a spin-orbit-coupled quantum ring is investigated within linear response theory. We show that the finite width of the ring results in the appearance of Fano resonances in the conductance. This turns out to be a consequence of the spin-orbit interaction that leads to a breaking of the parity of the states localized in the ring. The resonances appear when the system is close to maxima of Aharonov-Casher conductance oscillations where spin transfer is heavily modified. When the spin-orbit coupling strength is detuned from the Aharonov-Casher maxima the resonances are broadened resulting in a dependence of the spin transport on the electron Fermi energy in contrast to predictions from one-dimensional models.
\end{abstract}

\maketitle
\section{introduction}
Electrical manipulation of spin polarization of carriers is one of the key elements for semiconductor spintronics devices. Since the proposal of the spin-field-effect transistor by Das and Datta\cite{sfet} particular attention is addressed to Rashba spin-orbit (SO) coupling.\cite{rashba} This interaction -- a relativistic consequence of the presence of the electric field within the device -- produces an effective magnetic field\cite{bso} for the moving electrons which makes their spins precess.
%This interaction can be used to control the spin of confined electrons.
The Rashba coupling has been successfully implemented in quantum devices that operate on an electron spin through the control of voltages applied to the electrodes in the system, such as, quantum gates\cite{nowack} and valves.\cite{qpc}

Proposals of spin-operating devices concern also spin-orbit-coupled quantum rings as a realization of universal quantum gates,\cite{foldi} spin beam splitters\cite{foldi2} or spin filters.\cite{molod} The electron transfer through quantum rings involve both the spin precession due to the SO interaction and the quantum interference effects related to Aharonov-Bohm\cite{fuhrer,ablor} and Aharonov-Casher effects.\cite{casher}
The latter spin-interference effect\cite{nitta1,frustaglia} results from the fact that the relative phase shifts for the wavefunction passing through both arms of the ring are spin dependent in the presence of SO interaction. The AC oscillations were probed experimentally in a HgTe single quantum ring,\cite{konig} in a single\cite{nitta2} and in an array of InGaAs quantum rings\cite{nitta3} or in an array of $\mathrm{Bi}_2\mathrm{Se}_3$ topological insulator quantum rings.\cite{qu}

Although the theoretical work on SO-coupled quantum rings is rich, it is based mostly on the idealized case of a ring with infinitesimal narrow channels, i.e. the one-dimensional approximation. This approach allowed to obtain analytical description of charge\cite{molnar} and spin\cite{foldi} transport through the ring as function of the electron Fermi energy and the Rashba SO coupling strength. Theoretical studies concerning two-dimensional channels showed however that for an accurate description of transport through the spin-orbit-coupled ring, the finite width of the channels cannot be neglected.\cite{souma,konig,frustaglia} This is mainly due to the fact that for a finite width ring the spin is no longer well defined. Nonetheless, full calculations are rather scarce. A ring with two-dimensional channels has been studied within a tight binding formalism in Ref. [\onlinecite{frustaglia}] or in the framework of scattering matrix method in Ref. [\onlinecite{wu}].
%showing the magnetoconductance of a single-mode transport and the discrepancy between one- and two-dimensional calculations of the AC oscillations.
Ref. [\onlinecite{souma}] investigated the transport within the multiband Landauer-Buttiker formalism. This was later extended to describe the experimental data obtained in the presence of an external magnetic field.\cite{konig}

In this work we developed a novel calculation scheme that allows to study spin transport through a SO-coupled ring with two-dimensional channels. We show that the finite width of the channels along with SO interaction results in the appearance of Fano resonances in the conductance around the AC oscillations maxima. Those sharp peak/dip structures has been previously studied in the presence of an external magnetic field, where they were the result of the broken symmetry of states localized in the ring,\cite{poniedzia} in systems of quantum ring coupled to a quantum dot\cite{kobayashi} or in one-dimensional quantum rings containing impurities\cite{bellucci} and magnetic structures.\cite{faizabadi} Here we find that the Fano resonances originate from the coupling of the transmitted electron with the resonance states localized in the ring that have broken parity as a consequence of the SO interaction.

We find that in the resonances region the spin transport through the ring is strongly modified. We argue that the modification is caused by the coupling of the electron spin with the spin of the resonance states which is revealed by the application of an external magnetic field. When the SO coupling  strength is detuned from the AC oscillations maxima the Fano resonances are broadened which results in a dependence of the spin transport on the electron Fermi energy. This particular result was not present in previous studies on spin transformations in one-dimensional rings\cite{foldi} and is of importance for spintronics devices based on SO-coupled rings.

\section{Theory}
\subsection{System}
\begin{figure}[ht!]
\epsfxsize=70mm
                \epsfbox[16 16 570 821] {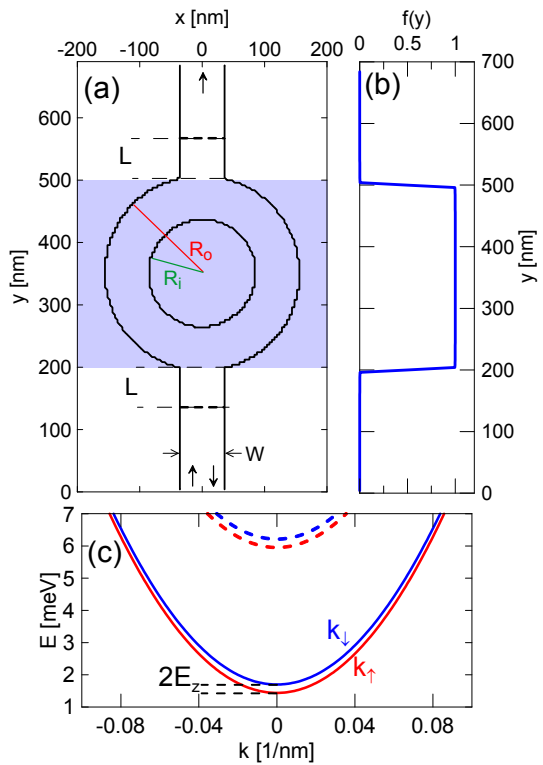}
                 \caption{(color online) (a) The contour of the confinement potential of the ring and the leads shown by the black curves. The region where Rashba coupling is present is marked with blue color. The dashed thick lines in the leads present the closed system of leads with length $L$ used to obtain the energies of the localized states (see text). (b) Function $f(y)$ that controls the spatial presence of the Rashba coupling. (c) Dispersion relation in the lead obtained for $B=0.5$ T.}
 \label{potential}
\end{figure}

We consider a system described by the effective mass Hamiltonian,
\begin{equation}
H=\left(\frac{\hbar^2\textbf{k}^2}{2m^*}+V_c(\textbf{r})\right)\textbf{1} + \frac{1}{2}g\mu_B B\sigma_z + H_{SIA},
\label{ham}
\end{equation}
where $V_c(\textbf{r})$ defines the confinement potential of the ring (with outer radius $R_o=152$ nm, inner radius $R_i=88$ nm and mean radius $R=120$ nm) and the leads, both with channel width $W=64$ nm. We adopt hardwall potential with $V_c=0$ inside the channels and $V_c=200$ meV outside (effectively an infinite barrier). The contour of the confinement potential is depicted in Fig. \ref{potential}(a) by the black curve.

The kinetic operator is $\textbf{k}=-i\nabla+\frac{e\textbf{A}}{\hbar}$. We include magnetic field $B$ directed perpendicular to the plain of the device. We choose the Lorentz gauge $\mathbf{A}=(A_x,A_y,0)=(0,Bx,0)$.

We include Rashba SO interaction with the Hamiltonian, $H_{SIA} = \alpha \nabla V\cdot (\sigma \times \mathbf{k})$ resulting from the electric field felt by the propagating electron. In order to allow for a well defined spin in the leads we assume that the Rashba coupling is present solely in the ring area by applying the electric field only therein. Experimentally this is realized\cite{nitta2} by applying a voltage between the substrate and a gate that is restricted to the ring area.

In the considered two-dimensional system we describe the potential that enters the Rashba Hamiltonian $H_{SIA}$ by $V(x,y,z) = V_{c}(x,y) + |e|F_z z f(y)$. We include inhomogeneous electric field $F_z(y) \equiv F_zf(y)$ that controls the coupling strength. In Fig. \ref{potential}(a) in the blue region the electric field is approximately equal $F_z$. The function that controls $F_z(y)$ has the form $f(y)=1/\pi^2(\arctan(y-y_1) + \pi/2)(-\arctan(y-y_2)+\pi/2)$, where $y_1 = 200$ nm and $y_2 = 500$ nm -- see Fig. \ref{potential}(a). The function is nearly step-like. We depict $f(y)$ by the blue curve in Fig. \ref{potential}(b). Finally, we obtain the Rashba operator for electric field in the growth direction that depends on the $y$-position (we neglect the derivatives of $V_c(\mathbf{r})$ as the wavefunction disappears in the proximity of the edges of the confinement potential):
\begin{equation}
\begin{split}
H_{SIA} = \alpha |e| F_z z \frac{\partial f(y)}{\partial y} (\sigma_z k_x - \sigma_x k_z) \\+  \alpha |e| F_z f(y) (\sigma_x k_y - \sigma_y k_x).
\end{split}
\end{equation}
We average over the $z$-direction assuming that the electron is in the ground-state for motion in the vertical excitation ($\langle z\cdot k_z \rangle=\frac{1}{2}i$), obtaining,
\begin{equation}
H_{SIA} = -\frac{i \alpha |e| F_z}{2}  \frac{\partial f(y)}{\partial y}\sigma_x + \alpha|e|  F_z f(y) (\sigma_x k_y - \sigma_y k_x).
\label{hsia}
\end{equation}
Note that the first term in the latter equation guarantees that the two-dimensional Rashba Hamiltonian is Hermitian when the strength of the coupling is nonhomogeneously varied along the $y$ direction. In this way we avoid artificial symmetrization needed in the previous work with nonhomogeneous Rashba coupling.\cite{palyi}

We employ material parameters for $\mathrm{In}_{0.5}\mathrm{Ga}_{0.5}\mathrm{As}$ alloy, i.e. $m^*=0.0465m_0$, $g=-8.97$. The Rashba coupling parameter $\alpha=0.572\;\mathrm{nm}^2$ is adopted from work Ref. [\onlinecite{alpha}].

\subsection{Method}
We solve the transport problem through the solution of the stationary Schr\"odinger equation $H \Psi = E \Psi$ assuming boundary conditions such that the electron enters the system from the bottom electrode and can either be backscattered or be transmitted through the system. The problem is solved on a grid with $\Delta x = \Delta y = 4$ nm using the finite difference approach, employing gauge-invariant discretization of the kinetic energy operator for each of the spinor $\Psi$ components:
\begin{equation}
\begin{split}
\frac{\hbar^2\textbf{k}^2}{2m^*}\Psi_{i,j}=&\\
&\frac{\hbar^2}{2m^*\Delta x^2}(4\Psi_{i,j}-C_y\Psi_{i,j-1}\\
&-C_y^*\Psi_{i,j+1}-C_x\Psi_{i-1,j}-C_x^*\Psi_{i+1,j}),
\end{split}
\end{equation}
where $\Psi_{i,j}=\Psi(x_i,y_j)$, $C_y=\exp[-i\frac{e}{\hbar}\Delta x A_y]=\exp[-i\frac{e}{\hbar}\Delta x B x]$, and $C_y=\exp[-i\frac{e}{\hbar}\Delta x A_x]=1$.
The derivatives in the SO Hamiltonian (\ref{hsia}) are discretized straightforward.

Hereafter we describe the applied method.

\subsubsection{Lead eigenstates}
We start by obtaining the asymptotic states in the leads far away from the ring (i.e. for $y=0$). In the leads: $\left[-i\hbar\frac{\partial}{\partial y},H\right]=0$ and the spinor of propagating wave can be written as
\begin{equation}
\Psi(x,y)=e^{iky}\left(
\begin{array}{c} \Psi_{\uparrow}^k(x)\\
\Psi_{\downarrow}^k(x)
\end{array}\right).
\end{equation}
By inserting this form of the spinor to the discretized Hamiltonian (\ref{ham}) we obtain an one-dimensional eigenproblem for transverse quantization in the lead.
We plot the dispersion relation in Fig. \ref{potential}(c). The energies of (split by the Zeeman energy $2E_z$) spin-up and spin-down states are plotted with the red and blue curves, respectively.

In the present work we consider the range of Fermi energies such $E_f$ lies below the energy of the third subband [the dashed red curve in Fig. \ref{potential}(c)]. In this case there are four possible values of the electron wavevector for a given electron Fermi energy. It can either belong to the lowest subband having $k_\uparrow$ or $-k_\uparrow$ and spin oriented along the $z$ direction or belong to the second subband having $k_\downarrow$ or $-k_\downarrow$ and spin oriented antiparallel to the $z$ direction.

Unless stated otherwise in the calculations we lift the spin degeneracy of the states in the leads by applying a residual magnetic field with $B=0.1$ mT, which does not induce any observable orbital effects.

\subsubsection{Boundary conditions}

The boundary conditions applied in the method assume that the electron enters the system with a given wavevector $k_{inc}$ (corresponding to a given energy) and can exit the system with the combination of positive wavevectors available for this given energy.

Let us first consider the output channel. The wave function in such channel is a combination of channel eigenstates with positive wavevectors (as we assume no backscattered waves in the output lead). Let us add to the derivative:
\begin{equation}
\frac{\partial\Psi(x,y)}{\partial y}=\sum_{k>0} ikc_k^{\mathrm{out}} \exp[iky]\left(
\begin{array}{c} \Psi_{\uparrow}^k(x)\\
\Psi_{\downarrow}^k(x)
\end{array}\right),
\end{equation}
$ik_{inc}\Psi(x,y)$, obtaining:
\begin{equation}
\begin{split}
\frac{\partial\Psi(x,y)}{\partial y}=\sum_{k>0} i(k+k_{inc}) c_k^{\mathrm{out}} \exp[iky]\left(
\begin{array}{c} \Psi_{\uparrow}^k(x)\\
\Psi_{\downarrow}^k(x)
\end{array}\right)\\-ik_{inc}\Psi(x,y).
\end{split}
\end{equation}
From the discretized form of the derivative:
\begin{equation}
\frac{\partial\Psi(x,y)}{\partial y}=\frac{\Psi(x,y+\Delta y)-\Psi(x,y-\Delta y)}{2\Delta y}
\end{equation}
we obtain,
\begin{equation}
\begin{split}
\Psi(x,y+\Delta y)=&2\Delta y\sum_{k>0} i(k+k_{inc})c_k^{\mathrm{out}} \exp[iky]\left(
\begin{array}{c} \Psi_{\uparrow}^k(x)\\
\Psi_{\downarrow}^k(x)
\end{array}\right),\\
&+\Psi(x,y-\Delta y)-2\Delta y ik_{inc}\Psi(x,y).
\end{split}
\end{equation}

The same procedure leads to the form of the boundary condition in the bottom of the computational box. Only now the sum includes a positive wavevector of incoming electron $k_{inc}$ and the two backscattered waves with negative wavevector. For instance, for transport of electron with wavevector $k_\uparrow$:

\begin{equation}
\begin{split}
\Psi(x,y-\Delta y)=&-2\Delta y\sum_k^{k\neq k_\downarrow} i(k+k_{inc})c_k^{\mathrm{in}} \exp[iky]\left(
\begin{array}{c} \Psi_{\uparrow}^k(x)\\
\Psi_{\downarrow}^k(x)
\end{array}\right),\\
&+\Psi(x,y+\Delta y)+2\Delta y ik_{inc}\Psi(x,y).
\end{split}
\end{equation}

We use the above forms of the wave functions to obtain the boundary conditions, i.e. $\Psi(x,y+\Delta y)$ for the top edge of the computational box, and $\Psi(x,y-\Delta y)$ at the bottom of the computational box (at the left and right edge of the mesh we assume $\Psi=0$). The used boundary conditions are transparent, i. e. the transport results do not depend on the length of the leads.

\subsubsection{Solving the transport problem}

We solve the system of equations produced by the discretization of the Schr\"odinger equation with the boundary conditions described above.
In the present method on the one hand the amplitudes $c_k^{\mathrm{in}}$ and $c_k^{\mathrm{out}}$ are required for the boundary condition and on the other hand they can be obtained from the solution of the Schr\"odinger equation. Thus we assume a initial values of the amplitudes (namely $c_{k_{\uparrow}}^{\mathrm{in}}=c_{k_{\uparrow}}^{\mathrm{out}}=1$ for $k_{inc}=k_{\uparrow}$ -- however we checked that the particular choice of the initial values does not change the final result) and put them into the boundary conditions. Then we solve the Schr\"odinger equation. From the solution we extract new values of the amplitudes by projection (in the input and output lead) of the function,

\begin{equation}
\Psi(x,y)=\sum_k c_k \exp[iky]\left(
\begin{array}{c} \Psi_{\uparrow}^k(x)\\
\Psi_{\downarrow}^k(x)
\end{array}\right).
\label{aswf}
\end{equation}
(that accounts all possible wavevectors for a given energy) onto the solution and solve again the Schr\"odinger equation. Such procedure is repeated until convergence is reached -- the extracted amplitudes do not change in the subsequent iterations, and the amplitudes $c_k$ are such that in the input channel there is only one incoming wave and in the output lead there are no backscattered waves.

We calculate transport probability from the ratio of the probability currents $j_k$ in the leads for respective wave vectors:

\begin{equation}
T_{k\rightarrow k'}=\left| \frac{c_{k'}^{\mathrm{out}}}{c_k^{\mathrm{in}}}\right|^2\cdot\frac{j_{k'}}{j_{k}}.
\end{equation}

The conductance $G$ is calculated as a sum of the transmission probabilities over available subbands, i.e. $G=\frac{2e^2}{h}\sum_{i}^{k_\uparrow,k_\downarrow}\sum_{j}^{k_\uparrow,k_\downarrow}T_{i\rightarrow j}$.

Described approach allows one to study the electron transport for {\it a given Fermi energy} in contrast to the methods involving transmission of a wavepacket\cite{ablor} in which the packet is not monoenergetic -- i.e. it consist of components from a finite range of energies. Also as the approach is based on an exact solution of the Schr\"odinger equation it naturally includes evanescent modes that can appear in the ring.

\section{Results}
\subsection{Fano resonances}

 \begin{figure}[ht!]
\epsfxsize=85mm
                \epsfbox[50 207 560 630] {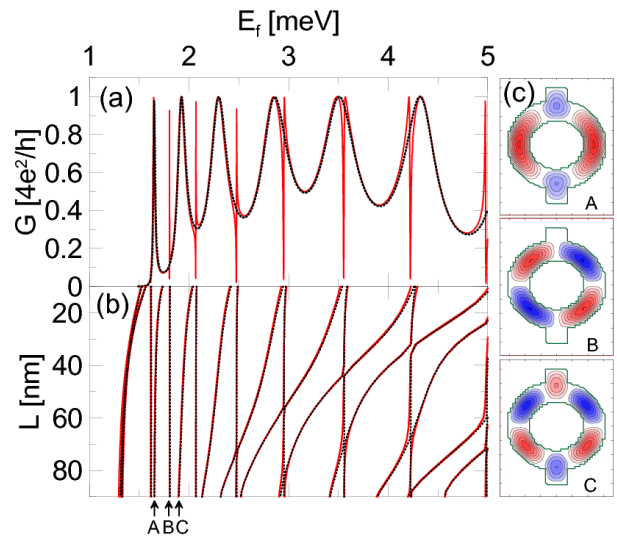}
                 \caption{(color online) (a) Conductance of the ring as function of the Fermi energy. (b) Energy spectrum for a closed system of a ring with leads of length $L$. In (a) and (b) black dotted curves represent results obtained for $F_z=0$ and red solid curves for $F_z=50\;\mathrm{kV/cm}$. (c) Real part of the spin-up wavefunction corresponding to the states A, B and C marked with blue arrows on (b) obtained without SO coupling for $L=80$ nm. Green contours depict the profile of the confinement potential.}
 \label{conductfz50}
\end{figure}

Let us start with the case of no SO coupling. In the ring there are two types of localized states -- states with positive and negative parity with respect to the $y=0$ axis. We inspect those states by diagonalization of Hamiltonian (\ref{ham}) for a closed system with varied length of the leads given by $L$ [see Fig. \ref{potential}(c)] and plot the energy spectrum with black dotted curves in Fig. \ref{conductfz50}(b). Note that in fact each curve corresponds to the energy of spin degenerate state. In Fig. \ref{conductfz50}(c) we plot the real part of the wavefunction of the spin-up states whose energies we mark by A, B and C. The A and C states are the states of positive parity and due to the fact that their wavefunction is nonzero in the leads their energy changes with $L$. On the other hand the wavefunction of negative-parity state B is zero in the leads and its energy is independent of $L$.

When the electron is transmitted through the ring the current carrying state from the lead (which is a state of positive parity, i.e. the ground-state of transverse excitation) couples to the localized states with positive parity. The conductance of the ring as function of electron Fermi energy [see black dotted curve in Fig. \ref{conductfz50}(a)] exhibits wide resonances due to this coupling. In the absence of SO interaction the localized states of the negative parity
are {\it bound}, i.e. their lifetime is infinite in spite of the fact that their energy lies in the energy continuum -- above the lowest subband transport threshold. On the other hand the energy of those states is still below the transport threshold for the second subband -- with wave functions of negative parity with respect to the axis of the channel.

When the SO coupling is introduced (we discuss first the case of weak Rashba coupling with $F_z=50\;\mathrm{kV/cm}$) the parity of the localized states is no longer well defined. For instance mean values of parity operator for states A, B, C are $0.939, -0.962, 0.957$ respectively. Due to the broken symmetry, the current carrying state from the lead couples now to all the localized states. This results in the appearance of sharp peaks in the conductance, plotted by the red curve in Fig. \ref{conductfz50}(a), in addition to the wide resonances. These sharp peaks are Fano resonances with characteristic asymmetric dip/peak structures. Their energy corresponds to the energy of states localized purely in the ring. The small width of the resonances is reflected in a finite but long lifetime of the resonance state. In Fig. \ref{conductfz50}(b) the red curves present the energy spectrum of a closed system as function of $L$ in the presence of SO coupling. Note that now due to the fact that the states lack a well defined parity there appear anticrossings in the spectrum.

\begin{figure}[ht!]
\epsfxsize=85mm
                \epsfbox[14 225 575 626] {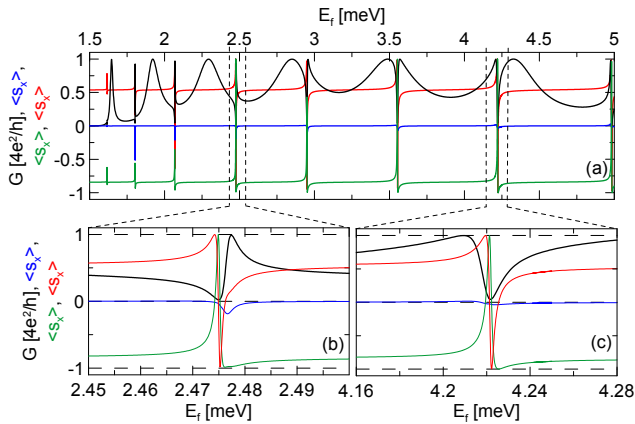}
                 \caption{(color online) (a) Conductance through the ring (black curve) and mean spin components (red and green solid curves) at the output lead for the electron with initial spin polarization along the $z$-direction.
                 %Dashed curves present spin components at the output multiplied by the conductance G.
                 (b, c) Close-ups of the resonances in plot (a).}
 \label{spinfz50}
\end{figure}

\subsection{Spin transport}
Let us now inspect spin transport through the ring. We consider the transmission of the electron with spin initially polarized along the $z$-direction (from the $k_{\uparrow}$ subband) and study the spin state at the output of the system.\cite{spin} However for the considered residual magnetic field $B=0.1$ mT the transfer probabilities from both subbands are exactly the same and the spin at the output of the ring is exactly opposite.

In Fig. \ref{spinfz50} we plot the conductance (black curve) and mean spin components at the output lead by solid colored curves. Notice that outside the resonance regions the output spin orientation remains unchanged when $E_f$ is varied -- see the red and green curves -- which is in agreement with the results of Ref. [\onlinecite{foldi}]. However when the electron Fermi energy is tuned to a resonance value the spin at the output is modified -- as seen clearly in Figs. \ref{spinfz50}(b, c).

\begin{figure}[ht!]
\epsfxsize=85mm
                \epsfbox[21 215 580 650] {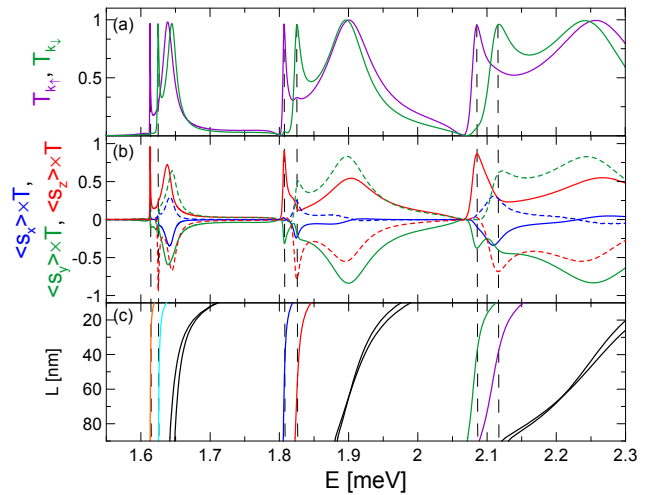}
                 \caption{(color online) (a) Transfer probability for electron incoming with spin polarized parallel (purple curve) and antiparallel (green curve) to the $z$ direction. (b) Mean values of the spin components in the $x,y$ and $z$ directions at the end of output lead multiplied by the transfer probabilities. Solid curves are obtained for transport from $k_{\uparrow}$, dashed curves for $k_{\downarrow}$. (c) Eigenenergies of closed system with leads of length $L$. The results are obtained for $B=0.02$ T.}
 \label{fz50b002}
\end{figure}

In order to further explore the spin changes in the narrow resonance regions let us increase the magnetic field, such that spin degeneracy of the localized resonance states is lifted, namely we apply $B=0.02$ T. Now the transfer probabilities for electron incoming from the subband $k_{\uparrow}$ and $k_{\downarrow}$ are no longer the same. In Fig. \ref{fz50b002}(a) with the purple curve we plot the transfer probability for an electron incoming from the lowest subband in the lead, i.e. $k_{\uparrow}$ with spin polarized parallel to the $z$ direction. With the green curve we show the transfer probability for an electron incoming from the second subband, i.e. $k_{\downarrow}$  with spin polarized antiparallel to the $z$ direction. We find that the Fano resonances from Figs. \ref{conductfz50}(a) and \ref{spinfz50}(a) are now converted to sharp peaks in the transfer probabilities occurring separately in both $T_{k_{\uparrow}}$ and $T_{k_{\downarrow}}$. In Fig. \ref{fz50b002}(c) we plot the energy spectrum of the closed system with varied length of the leads. Notice that the magnetic field splits the spin doublets of the localized states. The states have nonzero average spin component in the $z$ direction (with $|\langle s_z \rangle| \simeq 0.9$). The states lying lower in energy have $\langle s_z \rangle < 0$, the states with higher in energies have $\langle s_z \rangle > 0$. The splitting energy of the doublet is not equal to the Zeeman splitting as the spin and orbital parts of the wavefunction of the states are mixed by the Rashba coupling present in the ring. Namely the energy of twice the Zeeman splitting is $2E_z = |g\mu_BB| = 10\;\mu\mathrm{eV}$ and the energy difference between the states whose energies we mark with orange and light blue curves is $12\;\mu\mathrm{eV}$ and for the pair plotted with blue and red curves the energy difference is $17\;\mu\mathrm{eV}$. The mean values of the spin operators in the $x$ and $y$ direction are zero.

The peaks in the transport probabilities for an electron with initially spin polarized parallel to the $z$-direction [see purple curve in Fig. \ref{fz50b002}(a)] appear for energies equal to those of the resonance states marked with orange, blue and green colors in Fig. \ref{fz50b002}(c) -- with positive $\langle s_z \rangle$. For opposite spin orientation peaks are present for energies corresponding to the second state from the spin doublet (with energies marked with light blue, red and purple curves) -- with negative $\langle s_z \rangle$. This indicates that resonances in the transfer probability appear when the spin of the localized state matches the orientation of the spin of the incoming electron.

Let us now inspect the average spin components at the output of the system. In Fig. \ref{fz50b002}(b) we present the mean $x, y$ and $z$ spin components multiplied by the conductance with blue, green and red curves respectively. Solid curves correspond to initial spin-up polarization and the dashed one to initial spin-down polarization. We observe that when the transported electron couples to the resonance states localized in the ring the spin at the output is close to the average spin of the resonance state (see the peaks marked with vertical dashed lines). Outside of the resonances we observe that the spin of the transferred electron deviates from the $z$ or $-z$ direction.

\subsection{Dependence of the spin orientation on the Fermi energy for increased SO strength}

\begin{figure}[ht!]
\epsfxsize=65mm
                \epsfbox[17 18 570 820] {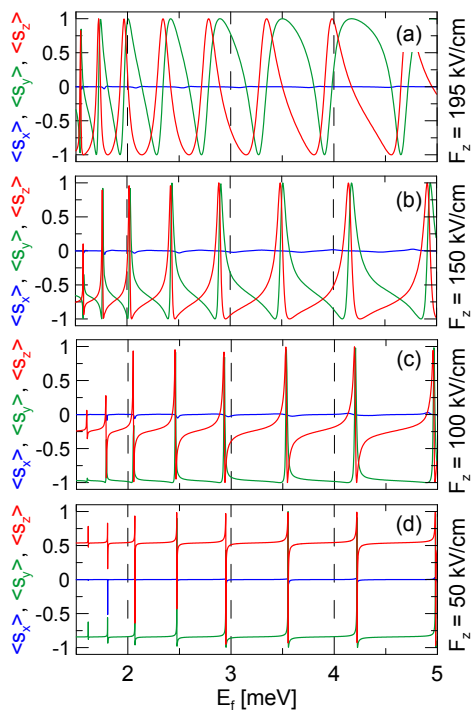}
                 \caption{(color online) Mean values of the spin components at the output lead obtained for different strengths of the Rashba coupling (as marked on the right side plots).}
 \label{spincross}
\end{figure}

Lets go back to the case of residual magnetic field $B=0.1$ mT and inspect the dependence of the spin transport on the electron Fermi energy for increased strength of the SO coupling. In Fig. \ref{spincross} we plot the mean spin components for four $F_z$ values. We observe distinct broadening of the spin changes at the Fano resonances. The broadening of the resonances corresponds to a reduction of the lifetime of the resonance states in the ring.
%{\it In fact in the eigenspectrum of the closed system we do not find any states whose energy would not depend on the lead width $L$ . As the $F_z$ is increased the SO coupling mix the states of the closed system such there are no states purely localized in the ring.}

\begin{figure}[ht!]
\epsfxsize=85mm
                \epsfbox[48 261 540 583] {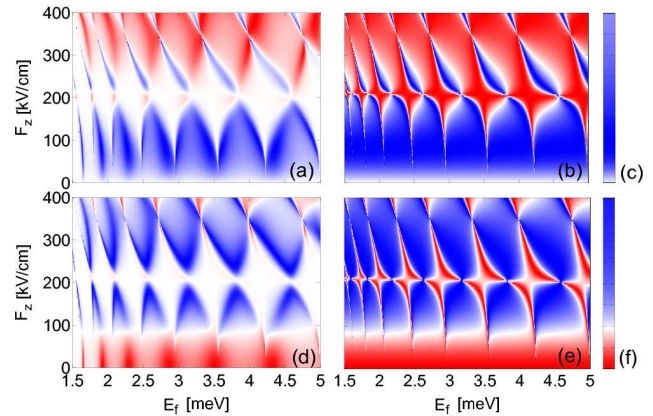}
                 \caption{(color online) Mean spin components [(a) -- $\langle s_y \rangle$, (d) -- $\langle s_z \rangle$] at the output of the system multiplied by the conductance as function of the Rashba coupling strength and the Fermi energy. Plots (b) and (e) shows mean spin components at the output of the ring $\langle s_y \rangle$ and $\langle s_z \rangle$ respectively. (c) and (d) are spin components independent of $E_f$ as calculated from the 1D approximation (see text).}
 \label{spin}
\end{figure}

Although the Fermi energy affects the spin evolution in the ring, the spin measured at the output (i.e. multiplied by the G) remains unchanged as function of $E_f$ for $F_z< 200\;\mathrm{kV/cm}$. This is made clear by the contour maps of the mean spin components multiplied by the conductance presented in Fig. \ref{spin}(a) for the $y$ component and in Fig. \ref{spin}(d) for the $z$ component. Outside the narrow Fano region spin changes are masked by the blocked transport through the ring.

Figures \ref{spin}(a,b) show that as the strength of the Rashba coupling is varied the conductance of the ring is changed due to the phase shift of the wavefunctions traveling in the left and right arm of the ring -- the AC effect that modifies the conductance by,\cite{nitta1}
\begin{equation}
G_{AC} = \frac{e^2}{h} \left[ 1 - \cos \left( \pi \sqrt{1+\left( \frac{2Rm^*\alpha|e|F_z}{\hbar^2}\right)^2}  \right) \right].
\label{aceq}
\end{equation}
For appropriately chosen strength of the SO coupling transport through the ring is quenched.\cite{molnar, souma} For parameters taken in the present calculation the first AC oscillations minimum ($G_{AC}=0$) is present around $F_z = 200\; \mathrm{kV/cm}$ which can be observed in Figs. \ref{spin}(a,d). On the other hand the first maxima of the AC oscillations appear for $F_z=0$ and around $F_z=340\; \mathrm{kV/cm}$ (note that the quenching of the conductance is found at slightly lower values of $F_z$ for higher Fermi energy).

%Moreover  for $F_z < 200\;\mathrm{kV/cm}$ we observe that spin evolution can be explained simply by spin precession in the SO effective magnetic field as if the electron was moving in a straight channel -- namely for $F_z = 90\;\mathrm{kV/cm}$ for which the $2R=\lambda_{SO}/2=(\pi\hbar^2)/(4|e|F_z\alpha m^*)$ the spin at the output of the ring is after half rotation around $x$ direction -- it is directed antiparallel to the $y$ direction.

Similarly to the dependence of the spin on the electron Fermi energy observed before the first AC oscillations minimum ($F_z<200\;\mathrm{kV/cm}$) we find that the spin changes for $F_z>200\;\mathrm{kV/cm}$ -- after the first AC oscillations minimum [see Figs. \ref{spin}(b,e)]. Similarly to the spin dependence in Fig. \ref{spincross}, they originate in the Fano resonances. Only now the resonances are associated with the second AC oscillations maximum around $F_z=350\;\mathrm{kV/cm}$. Moreover now the dependence of the spin on the Fermi energy is no longer masked by the quenched transfer probabilities and it is visible in the contour maps of Figs. \ref{spin}(a,d)

We conclude that the changes in the spin orientation originate at the Fano resonances appearing around/in the AC oscillations maxima and are broadened for SO coupling strength detuned from the AC oscillations maxima.

\subsection{Comparison with one-dimensional model}

In Ref. [\onlinecite{foldi}] it was found that when the electron is transferred through an one-dimensional spin-orbit-coupled ring its spin precesses around the $y$-direction by an angle $2\theta=2\arctan\left( -F_z|e|\alpha 2 m^* R/\hbar^2\right)$ which is independent of the electron Fermi energy. We calculated the spin after the rotation (taking R as the external radius of the ring -- the area where the SO is present) about the angle $2\theta$ and plot the spin components in Figs. \ref{spin} (c,f). We observe that before the first AC oscillations minimum both the present results: of the two-dimensional model and the results for the one-dimensional ring are similar with the exception of the spin resonances that are present in the first case. However as the SO coupling strength is increased -- the entanglement of the orbital and spin part of the wavefunction present for the system with finite width channel increases -- the discrepancy between our calculation and the results of the one-dimensional model increases. Namely, we find positive values of the $y$-spin component which is not seen in the one-dimensional approximation (note that the predictions of Eq. (\ref{aceq}) for AC oscillations maxima and minima still holds for this strength of SO coupling).

\subsection{Impact of the channel width and the ring radius}

\begin{figure}[ht!]
\epsfxsize=80mm
                \epsfbox[20 117 574 720] {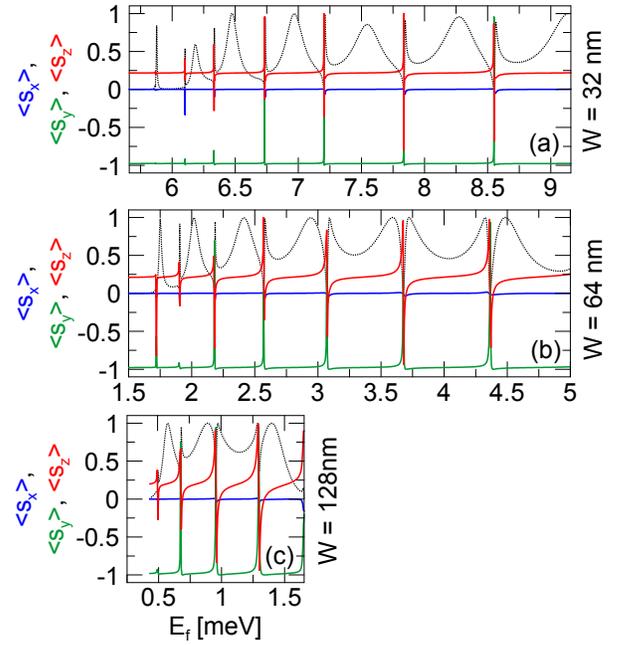}
                 \caption{(color online) The mean spin components at the output ($\langle s_x \rangle$ with blue, $\langle s_y \rangle$ with green and $\langle s_z \rangle$ with red curve) for $F_z=70\;\mathrm{kV/cm}$ and varied channel width $W$. Dotted black curves present the transport probabilities.}
 \label{wider}
\end{figure}

Let us now inspect the influence of the channel width on the spin behavior. We perform calculations in which we keep the mean radius of the ring constant and modify the channel width $W$. We observe that the dependence of the spin orientation on the Fermi energy is changed. In Fig. \ref{wider} we plot the spin components at the output of the ring for different channel width. The wide resonances of the spin at the output as function of Fermi energy found for the wide-channel ring [see Fig. \ref{wider}(c)] are transformed to sharp peaks for a ring with narrow channels [see Fig. \ref{wider}(a)]. As the strength of the Rashba coupling and the mean radius of the ring are kept constant, we conclude that the dependence of the spin on the electron Fermi energy is an effect of the two-dimensional character of the channels which gets weaker for decreased $W$, i.e. the resonances become narrower. This is in agreement with intuition that for infinitesimal narrow channels the dependence of $E_f$ should vanish (the peaks should be infinitesimally narrow) and the spin changes should not depend on the electron Fermi energy as found in the one-dimensional model. Note that in Fig. \ref{wider}(c) we plot only the Fermi energy range below the energy of the third subband in the lead.

\begin{figure}[ht!]
\epsfxsize=85mm
                \epsfbox[53 273 542 570] {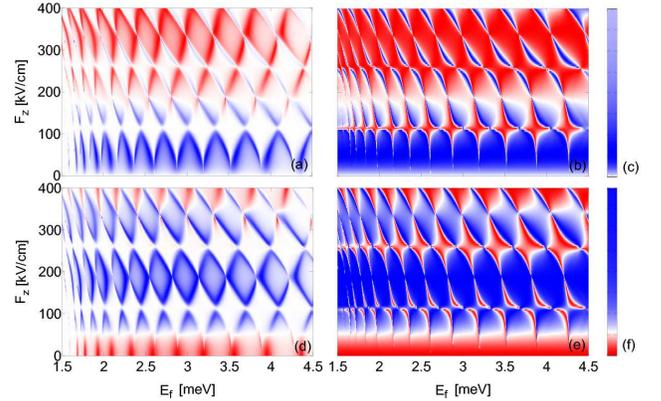}
                 \caption{(color online) Mean spin components [(a) -- $\langle s_y \rangle$, (d) -- $\langle s_z \rangle$] at the output of the system multiplied by the conductance as function of the Rashba coupling strength and the Fermi energy. Plots (b) and (e) show the mean spin components at the output of the ring $\langle s_y \rangle$ and $\langle s_z \rangle$ respectively. (c) and (d) are the spin components as calculated from the one-dimensional approximation (see text). Results are for a ring with radius $R=220$ nm.}
 \label{big}
\end{figure}

As expected from Eq. (\ref{aceq}) the position of the AC oscillations minima/maxima with respect to the Rashba coupling strength is inversely proportional to the ring radius. In Fig. \ref{big} we the present results obtained for a ring with mean radius $R=220$ nm. Firstly, we observe that spin components oscillate more frequently as function of the Fermi energy which can be attributed to the lower spacings between the energies of the localized states of a larger ring. Nevertheless, we observe a similar qualitative spin behavior with respect to the Fermi energy as in the case of a smaller ring [compare Figs. \ref{big}(b,e) with Figs. \ref{spin}(b,e)]. Only now we are able to observe more periods of AC oscillations. We again find that the spin changes originate from the Fano resonances, which appear when the SO strength is tuned to the AC conductance oscillation maxima -- i.e. for $F_z \simeq 185\; \mathrm{kV/cm}$ and $F_z \simeq 320\; \mathrm{kV/cm}$. However, in terms of the mean spin values multiplied by the conductance, we observe a dependence on the Fermi energy only for strong SO coupling (namely $F_z> 200\; \mathrm{kV/cm}$) similarly to the case of a smaller ring. Also the correspondence with the one-dimensional results [see Figs. \ref{big}(c,f)] remains the same as for smaller ring radius.

\section{Summary and conclusions}
In summary, we have studied the spin and charge transport through a spin-orbit-coupled quantum ring with two-dimensional channels. We found Fano resonances of the conductance which are present for non-zero SO coupling strength tuned to the maxima of AC oscillations. This narrow resonances are an effect of coupling of the current carrying states from the leads with the localized states in the ring that have a long lifetime. The coupling is possible due to the breaking of the parity of those states by the SO interaction.

In the Fano resonances spin transport through the ring is modified. We argue that the modification is due to the coupling of the transferred electron spin with the spin of the resonance states which we investigated in the presence of finite external magnetic field.

When the SO coupling strength is such that system is outside the AC oscillations maxima the Fano resonances are broadened. In that case the spin modification is translated into a wide dependence of the spin transport on the Fermi energy. The latter result is in contrast to the findings of the one-dimensional model\cite{foldi} which employed spin transformations (independent of $E_f$) performed in a quantum ring to realize a universal set of quantum gates. However when the width of the channels is decreased the resonances that results into the dependence of spin transport on the Fermi energy become narrower -- the results tend to the prediction of the one-dimensional model.

Moreover by the direct comparison of the results of the one-dimensional model and the two-dimensional calculation we found that for strong SO coupling the spin evolution proves to behave in a way exceeding the predictions for an one-dimensional ring even outside the resonances region.

\section*{Acknowledgements}
This work was supported by the "Krakow Interdisciplinary PhD-Project in Nanoscience and Advanced Nanostructures" operated within the Foundation for Polish Science MPD Programme co-financed by the EU European Regional Development Fund, the Project No. N N202103938 supported by Ministry of Science an Higher Education
(MNiSW) for 2010–2013, the Belgian Science Policy (IAP) and the Flemish Science Foundation (FWO-V1). Calculations were performed in ACK\---CY\-F\-RO\-NET\---AGH on the RackServer Zeus.


\begin{thebibliography}{00}
\bibitem{spin} Mean values of the $x, y$ and $z$ spin components are calculated on a normalized wavefunction on a stripe at the end of the output lead.
\end{thebibliography}
\end{document}